\documentclass{ws-mplb}
\usepackage{graphicx}

\begin{document}

\markboth{A.~Sherman}{Strong Coupling Diagram Technique for the Three-Band Hubbard Model}

%
\catchline{}{}{}{}{}
%
\title{Strong Coupling Diagram Technique\\for the Three-Band Hubbard Model}

\author{A.~SHERMAN}

\address{Institute of Physics, University of Tartu, Ravila 14c, 50411 Tartu, Estonia}

\maketitle

\begin{history}
\received{(Day Month Year)}
\revised{(Day Month Year)}
\end{history}

\begin{abstract}
Using the strong coupling diagram technique equations are derived for hole Green's functions of the three-band Hubbard model, which describes Cu-O planes of high-$T_c$ cuprates. The equations are self-consistently solved in the approximation, in which the series for the irreducible part in powers of the oxygen-copper hopping constant is truncated to two lowest-order terms. For parameters used for hole-doped cuprates the calculated energy spectrum consists of lower and upper Hubbard subbands of predominantly copper nature, oxygen bands with a small admixture of copper states and the Zhang-Rice states of mixed nature, which are located between the lower Hubbard subband and oxygen bands. The spectrum contains also pseudogaps near transition frequencies of Hubbard atoms on copper sites.
\end{abstract}


\section{Introduction}
The three-band Hubbard model, named also the Emery model,\cite{Emery} contains a minimal set of orbitals, which is necessary for the description of Cu-O planes of cuprate high-$T_c$ superconductors -- the copper $d_{x^2+y^2}$ and oxygen $p_x$ and $p_y$ orbitals. Previously the model was considered by different methods including the dynamic cluster approximation,\cite{Macridin} dynamic mean-field approximation,\cite{Weber,Medici} variational cluster approach\cite{Arrigoni} and density-matrix-renormalization-group calculations.\cite{White} These works were mainly aimed at the investigation of manifestations of the Zhang-Rice states\cite{Zhang} in spectral functions and at the comparison of the low-frequency part of the spectrum with the spectrum of the one-band Hubbard model.

In this work we use the strong coupling diagram technique\cite{Vladimir,Metzner,Craco,Pairault,Sherman06} for investigating the Emery model. Recently it was shown\cite{Sherman15} that this approach is able to describe the Mott metal-insulator transition in the one-band Hubbard model. The obtained value of the critical repulsion is close (identical for the semi-elliptical initial density of states) to that obtained in the Hubbard-III approximation.\cite{Hubbard64} In addition, it was shown that spectral functions and the momentum distribution of the one-band Hubbard model, calculated in this theory, are in reasonable agreement with data of Monte Carlo simulations in a wide range of electron concentrations. The approach is computationally simple and, therefore, it would be of interest to extend it to multi-band generalizations of the Hubbard model, for which other approaches need in intensive computations. To our knowledge the present work is the first attempt to apply the strong coupling diagram technique to one of such models -- the Emery model.

In application to this model the method is based on the serial expansion in powers of the copper-oxygen hybridization term. The elements of the arising diagram technique are cumulants of hole creation and annihilation operators on copper sites and hopping lines proportional to unperturbed oxygen Green's functions. As in the diagram technique with the expansion in powers of an interaction, in the considered approach the linked-cluster theorem allows one to discard disconnected diagrams and to carry out partial summations in connected diagrams. The central element of equations obtained with this diagram technique is the copper irreducible part -- a sum of all irreducible diagrams, which cannot be separated into two disconnected parts by cutting a hopping line. In the below calculations an infinite series of diagrams forming the irreducible part is truncated to two lowest-order terms, and the equations are self-consistently solved for parameters adopted for hole doped cuprates. Obtained energy spectra contain two Hubbard subbands of predominantly copper nature, oxygen bands with an admixture of copper states and the Zhang-Rice states of mixed nature, which are located between the lower Hubbard subband and oxygen bands. Pseudogaps near transition frequencies of Hubbard atoms on copper sites are observed in the spectrum as well. The spectral intensity of the Zhang-Rice states and widths of the pseudogaps vary strongly with parameters. Most prominent features of the spectra are similar to those obtained by other, computationally more complicated, methods. In some cases, for close parameters there is agreement in locations, dispersions and relative intensities of bands. 

\section{Main formulas}
Taking into account phases of the copper $d_{x^2+y^2}$ and oxygen $p_x$ and $p_y$ orbitals, the Hamiltonian of the two-dimensional (2D) Emery model can be written as
\begin{eqnarray}\label{Hamiltonian}
H&=&\Delta\sum_{\bf lz\sigma}p^\dagger_{\bf l+z,\sigma}p_{\bf l+z,\sigma}-t_{pd} \sum_{\bf lz\sigma}\sum_{s=\pm 1}s\left(d^\dagger_{\bf l\sigma}p_{{\bf l}+s{\bf z},\sigma}+p^\dagger_{{\bf l}+s{\bf z},\sigma}d_{\bf l\sigma}\right) \nonumber\\
&&\quad -t_{pp}\sum_{\bf lz\sigma}\sum_{s,s'=\pm 1}ss'p^\dagger_{{\bf l}+s'{\bf z},\sigma}p_{{\bf l}+s\bar{{\bf z}},\sigma}+\frac{U}{2}\sum_{\bf l\sigma}n_{\bf l\sigma}n_{\bf l,-\sigma},
\end{eqnarray}
where $d^\dagger_{\bf l\sigma}$ creates a hole on a copper site {\bf l} with the spin projection $\sigma=\pm 1$; {\bf z} acquires the values {\bf x} and {\bf y}, which are vectors along the $x$ or $y$ axes with the length equal to a half of the copper intersite distance; $p^\dagger_{\bf l+z,\sigma}$ creates a hole on an oxygen site $\bf l+z$ located halfway between two copper sites; $\bar{{\bf z}}={\bf x}$ when ${\bf z=y}$ and $\bar{{\bf z}}={\bf y}$ when ${\bf z=x}$; $\Delta$ is the $d$-$p$ promotion energy; $t_{pd}$ and $t_{pp}$ are oxygen-copper and oxygen-oxygen hopping constants, respectively; $n_{\bf l\sigma}=d^\dagger_{\bf l\sigma}d_{\bf l\sigma}$ is the copper occupation number operators and $U$ is the copper on-site Coulomb repulsion. The Coulomb repulsion of holes on oxygen sites and between oxygen and copper sites can be taken into account on the mean-field level, since relevant hole occupations of oxygen sites are less than 30\%. Using values of the respective constants brought in literature, one can conclude that these interactions increase the effective value of $\Delta$ by $t_{pd}$ or less.

It is possible to get rid of phase factors $s$ in (\ref{Hamiltonian}). This can be done by passing to new operators
$$\widetilde{d}_{\bf l\sigma}={\rm e}^{i{\bf Ql}}d_{\bf l\sigma},\quad \widetilde{p}_{{\bf l}+s{\bf z},\sigma}=s\,{\rm e}^{i{\bf Ql}}p_{{\bf l}+s{\bf z},\sigma},\quad {\bf Q}=\left(\frac{\pi}{a},\frac{\pi}{a}\right).$$
Considering spectral functions one has to bear in mind that the above change shifts a hole momentum by the vector {\bf Q}.

We shall consider the hole Green's functions
$$G^d({\bf l'}\tau',{\bf l}\tau)=\langle{\cal T}\bar{d}_{\bf l'\sigma}(\tau')d_{\bf l\sigma}(\tau)\rangle,\quad
G^p({\bf l'+z'}\tau',{\bf l+z}\tau)=\langle{\cal T}\bar{p}_{\bf l'+z',\sigma}(\tau') p_{\bf l+z\sigma}(\tau)\rangle,$$
where angular brackets denote averaging over the grand canonical ensemble with the operator ${\cal H}=H-\mu\sum_{\bf l\sigma}d^\dagger_{\bf l\sigma}d_{\bf l\sigma}-\mu\sum_{\bf lz\sigma}p^\dagger_{\bf l+z,\sigma}p_{\bf l+z,\sigma}$, $\mu$ is the chemical potential,
$$d_{\bf l\sigma}(\tau)={\rm e}^{{\cal H}\tau}d_{\bf l\sigma}(\tau){\rm e}^{-{\cal H}\tau},\quad \bar{d}_{\bf l\sigma}(\tau)={\rm e}^{{\cal H}\tau}d^\dagger_{\bf l\sigma}(\tau){\rm e}^{-{\cal H}\tau},$$
and ${\cal T}$ is the chronological operator, which arrange operators from right to left in ascending order of times $\tau$.

In the used strong coupling diagram technique the copper-oxygen hybridization, the second term on the right-hand side of (\ref{Hamiltonian}), is considered as the perturbation, while other terms of $H$ form the unperturbed Hamiltonian. Terms of the power expansions for the $G^d$ and $G^p$ contain products of cumulants, belonging to copper sites, and unperturbed oxygen Green's functions, which are generated by the first and third terms of the Hamiltonian (\ref{Hamiltonian}). The Fourier transform of these Green's functions reads
\begin{equation}\label{Gp0}
G^p_0({\bf z'zk},i\omega_m)=\frac{\delta_{\bf z'z}(i\omega_m-\Delta+\mu) -\delta_{\bf z'\bar{z}}t_{pp}\gamma_{\bf k}}{(i\omega_m-\Delta+\mu)^2-(t_{pp}\gamma_{\bf k})^2},
\end{equation}
where $\omega_m=(2m+1)\pi T$ is the Matsubara frequency with the temperature $T$, {\bf k} is the 2D wave vector, $\gamma_{\bf k}=4\sin({\bf kx})\sin({\bf ky})$ and the Kronecker delta $\delta_{\bf z'z}=1$ when ${\bf z'=z}$ and 0 in the opposite case. The terms of the series can be presented diagrammatically: on-site cumulants are depicted by circles, which are connected by arrowed hopping lines corresponding to $t_{pd}^2 G^p_0$.

The diagram with copper external ends is said to be irreducible if it cannot be divided into two disconnected parts by cutting one hopping line. The sum of all such diagrams is termed the irreducible part and denoted by $K({\bf l'}\tau',{\bf l}\tau)$. In terms of this quantity the equation for the Green's function on copper sites reads
\begin{eqnarray}\label{Gdl}
G^d({\bf l'}\tau',{\bf l}\tau)&=&K({\bf l'}\tau',{\bf l}\tau)+t_{pd}^2 \int\!\!\!\!\int_0^\beta d\tau_1 d\tau_2\sum_{{\bf l}_1{\bf z}_1s_1}\sum_{{\bf l}_2{\bf z}_2s_2}K({\bf l'}\tau',{\bf l}_2\tau_2)\nonumber\\
&&\times G^p_0({\bf l}_2+s_2{\bf z}_2,\tau_2;{\bf l}_1+s_1{\bf z}_1,\tau_1) G^d({\bf l}_1\tau_1,{\bf l}\tau),
\end{eqnarray}
where $\beta=1/T$. Equation (\ref{Gdl}) shows that the transfer of a hole between copper sites ${\bf l'}$ and ${\bf l}$ can occur either in a process described by one of diagrams forming $K({\bf l'}\tau',{\bf l}\tau)$ or with one or several intermediate transitions to oxygen sites, which are followed by reverse transitions to copper sites. Coming on an oxygen site, a hole can move along all possible trajectories lying on oxygen sites. All these possibilities are summed out in the Green's function $G^p_0$ in (\ref{Gdl}).

After the Fourier transformation Eq.~(\ref{Gdl}) reads
\begin{equation}\label{Gdk}
G^d({\bf k},i\omega_m)=\left\{\left[K({\bf k},i\omega_m)\right]^{-1}-t^2_{pd} G^p_{0s}({\bf k},i\omega_m)\right\}^{-1},
\end{equation}
where $K({\bf k},i\omega_m)$ is the Fourier transform of $K({\bf l'}\tau',{\bf l}\tau)$ and $G^p_{0s}({\bf k},i\omega_m)$ is the Green's function for the symmetric, in accord with the hybridization term in (\ref{Hamiltonian}), combination of four oxygen orbitals around a copper site,
\begin{eqnarray}\label{Gp0s}
G^p_{0s}({\bf k},i\omega_m)&=&4\sum_{\bf zz'}\sin({\bf kz})\sin({\bf kz'})G^p_0({\bf z'zk},i\omega_m)\nonumber\\
&=&\frac{4\left[\sin^2({\bf kx})+\sin^2({\bf ky})\right](i\omega_m-\Delta+\mu)- 2t_{pp}\gamma^2_{\bf k}}{(i\omega_m-\Delta+\mu)^2-(t_{pp}\gamma_{\bf k})^2}.
\end{eqnarray}
Notice that the diagram series for the $G^d$ looks similar to that for the one-band Hubbard model,\cite{Sherman06} and even expressions for cumulants of different orders on copper sites are the same, since the same is the Hamiltonian, which determines the averaging and time dependencies of operators in the cumulants -- the last term in (\ref{Hamiltonian}). The only difference in the diagrams for $G^d$ and for Green's function of the one-band Hubbard model is in the meaning of the hopping lines. They correspond to $t_{pd}^2G^p_{0s}({\bf k},i\omega_m)$ in the former case and $t_{\bf k}$, the Fourier transform of hopping constants, in the latter.

In the same manner an equation for the hole Green's function on oxygen sites can be derived,
\begin{eqnarray}\label{Gpl}
G^p({\bf l'+z'},\tau';{\bf l+z},\tau)&=&G^p_0({\bf l'+z'},\tau';{\bf l+z},\tau)\nonumber\\
&&+t^2_{pd}\int \!\!\!\!\int^\beta_0 \!d\tau_1 d\tau_2\!\sum_{{\bf l}_1{\bf z}_1s_1}\sum_{{\bf l}_2{\bf z}_2s_2}\!s_1s_2 G^p_0({\bf l'+z'},\tau';{\bf l}_1+s_1{\bf z}_1,\tau_1)\nonumber\\
&&\times K({\bf l}_1\tau_1,{\bf l}_2\tau_2)G^p({\bf l}_2+s_2{\bf z}_2,\tau_2;{\bf l z},\tau),
\end{eqnarray}
which after the Fourier transformation reads
\begin{eqnarray}\label{Gpk}
G^p({\bf z'zk},i\omega_m)&=&G^p_0({\bf z'zk},i\omega_m)+4t^2_{pd}\sum_{{\bf z}_1{\bf z}_2}\sin({\bf kz}_1)\sin({\bf kz}_2)G^p_0({\bf z'z}_1{\bf k},i\omega_m)\nonumber\\
&&\times K({\bf k},i\omega_m)G^p({\bf z}_2{\bf zk},i\omega_m).
\end{eqnarray}
This equation can be solved by multiplying it by $\sin({\bf kz'})$ and summing over ${\bf z'}$. Denoting
\begin{eqnarray}\label{Gp0h}
G^p_{0h}({\bf zk},i\omega_m)&=&\sum_{\bf z'}\sin({\bf kz'})G^p_{0}({\bf z'zk},i\omega_m)\nonumber\\
&=& \frac{\sin({\bf kz})(i\omega_m-\Delta+\mu)-\sin({\bf k\bar{z}})t_{pp}\gamma_{\bf k}}{(i\omega_m-\Delta+\mu)^2-(t_{pp}\gamma_{\bf k})^2}
\end{eqnarray}
we get
\begin{equation}\label{Gpk2}
G^p({\bf z'zk},i\omega_m)=G^p_0({\bf z'zk},i\omega_m)+\frac{4t^2_{pd}G^p_{0h}({\bf z'k},i\omega_m)G^p_{0h}({\bf zk},i\omega_m)}{\left[K({\bf k},i\omega_m)\right]^{-1}-t^2_{pd} G^p_{0s}({\bf k},i\omega_m)}.
\end{equation}

\begin{figure}[t]
\centerline{\resizebox{0.95\columnwidth}{!}{\includegraphics{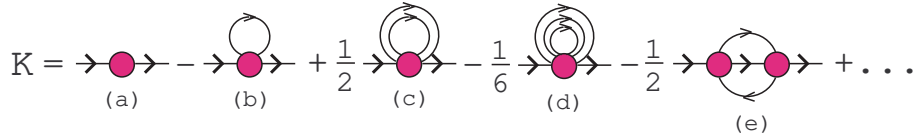}}}
\caption{Diagrams of the first four orders in $K({\bf k},i\omega_m)$.} \label{Fig1}
\end{figure}
Thus having an approximation for the irreducible part $K({\bf k},i\omega_m)$ one can calculate Green's functions (\ref{Gdk}) and (\ref{Gpk2}) for given parameters of Hamiltonian (\ref{Hamiltonian}) and a chemical potential. Diagrams of the four lowest orders in $t_{pd}^2$ are shown in Fig.~\ref{Fig1}. In the below calculations $K({\bf k},i\omega_m)$ is approximated by the sum of the diagrams (a) and (b), in the latter of which we use the above-mentioned possibility of partial summation, replacing the bare internal hopping line by the dressed one,
\begin{equation}\label{hopping}
t^2_{pd}G^p_{0s}({\bf k},i\omega_m)\rightarrow 4t^2_{pd}\sum_{\bf zz'}\sin({\bf kz})\sin({\bf kz'})G^p({\bf zz'k},i\omega_m).
\end{equation}
Expression for the cumulants of the first and second orders, which appear in $K$, are essentially simplified for chemical potentials satisfying the conditions
\begin{equation}\label{conditions}
T\ll\mu, \quad T\ll U-\mu
\end{equation}
(general expressions for these cumulants can be found in Ref.~12). Inequalities (\ref{conditions}) determine the range of $\mu$, which is of primary interest for the case of hole doping. For such chemical potentials $K$ reads
\begin{eqnarray}\label{K}
K(i\omega_m)&=&\frac{i\omega_m+\mu+(s_1-1)U/2}{(i\omega_m+\mu)(i\omega_m +\mu-U)}-\frac{s_2}{2}\frac{(i\omega_m+\mu-U)^2+U(i\omega_m+\mu)}{(i \omega_m+\mu-U)^2 (i\omega_m+\mu)^2}\nonumber\\
&&+\frac{3}{4}\frac{U^2\varphi(i\omega_m)}{(i\omega_m+\mu-U)^2 (i\omega_m+\mu)^2},
\end{eqnarray}
where
\begin{equation}\label{phi}
\varphi(i\omega_m)=\frac{4t_{pd}^2}{N}\sum_{\bf kzz'}\sin({\bf kz})\sin({\bf kz'}) G^p({\bf zz'k},i\omega_m),
\end{equation}
\begin{eqnarray}\label{sim}
s_1&=&T\sum_m\frac{(i\omega_m+\mu-U)^2+U(i\omega_m+\mu)}{(i \omega_m+\mu-U)^2 (i\omega_m+\mu)^2}\:\varphi(i\omega_m),\nonumber\\[-1.5ex]
&&\\[-1.5ex]
s_2&=&-T\sum_m\frac{U}{(i\omega_m+\mu)(i\omega_m +\mu-U)}\: \varphi(i\omega_m),\nonumber
\end{eqnarray}
and $N$ is the number of copper sites. Notice that in the used approximation $K$ in (\ref{K}) does not depend on wave vector.

It is now convenient to change to real frequencies in the above equations. Using the spectral representations, Poison summation formulas and inequalities (\ref{conditions}) equations (\ref{sim}) can be rewritten as
\begin{eqnarray}\label{sre}
s_1&=&\int_{-\infty}^0\frac{\omega+\mu-2U}{U(\omega+\mu- U)^2}\,{\rm Im}\,\varphi(\omega)\frac{d\omega}{\pi} +\int^{\infty}_0\frac{\omega+\mu+U}{U(\omega+\mu)^2}\, {\rm Im}\,\varphi(\omega)\frac{d\omega}{\pi},\nonumber\\[-1ex]
&&\\[-1ex]
s_2&=&\int_{-\infty}^0\frac{1}{\omega+\mu- U}\,{\rm Im}\,\varphi(\omega)\frac{d\omega}{\pi} +\int^{\infty}_0\frac{1}{\omega+\mu}\, {\rm Im}\,\varphi(\omega)\frac{d\omega}{\pi}.\nonumber
\end{eqnarray}
An equation for $\varphi(\omega)$ is obtained from the definition (\ref{phi}) and (\ref{Gpk2}), in which the irreducible part is given by (\ref{K}). The result reads
\begin{eqnarray}\label{phisc}
\varphi(\omega)&=&\frac{1}{N}\sum_{\bf k} \bigg\{\frac{(\omega+\mu-\Delta)^2-(t_{pp}\gamma_{\bf k})^2}{t_{pd}^2\left\{4(\omega+\mu- \Delta)\left[\sin^2({\bf kx})+\sin^2({\bf ky})\right]-2t_{pp}\gamma_{\bf k}^2\right\}}\nonumber\\
&&\quad\quad\quad\quad-\frac{\omega+\mu+(s_1-1)U/2}{(\omega+\mu)(\omega+ \mu-U)}+\frac{s_2}{2}\frac{(\omega+\mu-U)^2+U(\omega+\mu)}{(\omega+ \mu)^2 (\omega+\mu-U)^2}\nonumber\\
&&\quad\quad\quad\quad-\frac{3}{4}\frac{U^2\varphi(\omega)}{(\omega+\mu)^2 (\omega+\mu-U)^2}\bigg\}^{-1}.
\end{eqnarray}
After calculating $\varphi(\omega)$, $s_1$ and $s_2$ from equations (\ref{sre}) and (\ref{phisc}) the irreducible part (\ref{K}) and Green's functions (\ref{Gdk}) and (\ref{Gpk2}) can be worked out.

Equation (\ref{phisc}) allows us to determine analytically the behavior of $\varphi(\omega)$ near $\omega=-\mu$ and $\omega=U-\mu$, the transfer frequencies between levels of the Hubbard atom on copper sites. For $s_2\neq 0$ the result reads
\begin{equation}\label{s2neq0}
\varphi(\omega\approx -\mu)\approx 2s_2^{-1}(\omega+\mu)^2,\quad \varphi(\omega\approx U-\mu)\approx 2s_2^{-1}(\omega+\mu-U)^2,
\end{equation}
while for $s_2=0$
\begin{eqnarray*}
&&\varphi(\omega\approx -\mu)\approx\frac{s_1-1}{3}\,(\omega+\mu)- i\sqrt{\frac{4}{3}-\frac{(s_1-1)^2}{9}}\,|\omega+\mu|, \\
&&\varphi(\omega\approx U-\mu)\approx -\frac{s_1-1}{3}\,(\omega+\mu-U)- i\sqrt{\frac{4}{3}-\frac{(s_1-1)^2}{9}}\,|\omega+\mu-U|.
\end{eqnarray*}
We found that for the considered parameters the condition $s_2=0$ is fulfilled only for some $\mu$, which is deep in oxygen bands and corresponds to a heavily overdoped case. This case will not be considered here. For other $\mu$ equation (\ref{s2neq0}) is valid, which shows that $\varphi(\omega)$ and with it $K(\omega)$ (\ref{K}) are purely real in the vicinity of $\omega=-\mu$ and $\omega=U-\mu$. Therefore, the spectral functions $A({\bf k}\omega)= -{\rm Im}\,G({\bf k}\omega)/\pi$ and densities of states (DOS) $\rho(\omega)=N^{-1}\sum_{\bf k}A({\bf k}\omega)$ would have gaps near these frequencies if it were not for oxygen state damping, which transforms them into pseudogaps. Analogous, however much narrower, gaps at these frequencies are observed also in the one-band Hubbard model.\cite{Sherman15} Their appearance is a pathway to circumvent energetically unfavorable states with double occupancy of a site by holes or electrons. We suppose that the pseudogaps retain when terms of higher orders are included into consideration. Indeed, the function $\varphi(\omega)$ appears due to hopping loops in the local terms of $K$, and only this function may have an imaginary part in these terms. A term of the next order contains an additional loop, i.e. an additional multiplier $\varphi(\omega)$, and the extra multipliers $(\omega+\mu)^{-1}(\omega+\mu-U)^{-1}$ in a higher-order cumulant. Due to (\ref{s2neq0}) its contribution vanishes near $\omega=-\mu$ and $U-\mu$, and the contribution of the diagram (b) in Fig.~\ref{Fig1} dominates in these frequency ranges.

\section{Results and discussion}
Results of this section were obtained by the self-consistent solution of equations (\ref{sre}) and (\ref{phisc}), that was carried out by iteration. As the starting value for $\varphi(\omega)$ a ``Hubbard-I'' approximation of (\ref{phisc}) was used, in which $s_1$, $s_2$ and $\varphi(\omega)$ under the sum sign were set to zero and a small imaginary part of the order of $0.02t_{pd}$ was added to $\omega$ there. Moderate changes of an initial $\varphi(\omega)$ lead to the same converged function, and a variation of the artificial broadening influences only intensities of tails and depths of the pseudogaps (since in the above formulas $G^p_0$, $G^p_{0s}$ and $G^p_{0h}$ have no imaginary parts for real frequencies, in all stages of calculations $\omega$ in them was substituted with $\omega+i\eta$ with the above-mentioned value of $\eta$). From the obtained ${\rm Im}\,\varphi(\omega)$ and equation (\ref{K}) the function ${\rm Im}\,K(\omega)$ can be calculated. As follows from (\ref{Gdk}) and (\ref{Gp0s}), $K(\omega)$ coincides with $G^d({\bf k}\omega)$ for ${\bf k}=(0,0)$. Therefore, $K(\omega)$ has to satisfy the normalization condition
\begin{equation}\label{norm}
\int_{-\infty}^\infty {\rm Im}\,K(\omega)\,d\omega=-\pi
\end{equation}
and the Kramers-Kronig relation
\begin{equation}\label{KKr}
{\rm Re}\,K(\omega)={\cal P}\int_{-\infty}^\infty \frac{{\rm Im}\,K(\omega')}{\omega'- \omega}\frac{d\omega}{\pi},
\end{equation}
where ${\cal P}$ denotes the Cauchy principal value. We scaled the obtained ${\rm Im}\,K(\omega)$ to fulfill (\ref{norm}) and used (\ref{KKr}) for calculating ${\rm Re}\,K(\omega)$ from such obtained imaginary part, thereby ensuring correct analytic properties of the retarded Green's function. Spectral functions calculated from (\ref{Gdk}) and (\ref{Gpk2}) with this $K(\omega)$ were proved to satisfy the normalization conditions with good accuracy. Parameters chosen for calculations -- $U=8-12$, $\Delta=2-4$, $t_{pp}=0.5$ -- are close to those used for the Emery model earlier.\cite{Macridin,Weber,Medici,Arrigoni,White} Here and hereafter $t_{pd}\approx 1$~eV is set as the unit of energy and the distance between copper site as the unit of length.

\begin{figure}[tbh]
\centerline{\resizebox{0.7\columnwidth}{!}{\includegraphics{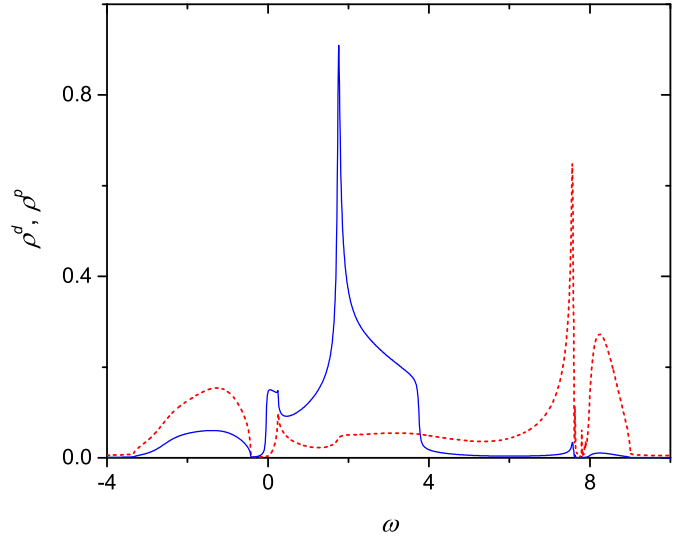}}}
\caption{Normalized to unity densities of copper (red dashed line) and oxygen (blue solid line) states for parameters $U=8$, $\Delta=2$ and $t_{pp}=0.5$. For $T=0$ the hole concentration $n=1.14$, of which 0.63 is related to copper sites and 0.25 on each of the two types of oxygen sites. The zero frequency corresponds to the Fermi level.} \label{Fig2}
\end{figure}
An example of the obtained results is shown in Fig.~\ref{Fig2}. Here copper and oxygen densities of states, $\rho^d$ and $\rho^p$, and hole concentrations are calculated from the respective Green's functions $G^d({\bf k}\omega)$ and $G^p({\bf xxk}\omega)$. The DOS obtained from $G^p({\bf yyk}\omega)$ is the same as shown in the figure. As seen from it, the spectrum consists of the lower Hubbard subband centered around $\omega=-2$, the upper Hubbard subband near $\omega=8$ and oxygen bands near $\omega=2$. If the upper subband is of predominantly copper nature, the lower subband has a considerable admixture of oxygen states. The pseudogaps mentioned in the previous section are seen around $\omega=0$ and 8. Near them sharp maxima of the DOS are observed. The lower one can be related to the Zhang-Rice states, since copper and oxygen contributions are comparable in them, as seen in Fig.~\ref{Fig2}. Besides, from equations of the previous section one can see that the location of these states are connected with the location of oxygen levels.

\begin{figure}[tbh]
\centerline{\resizebox{0.7\columnwidth}{!}{\includegraphics{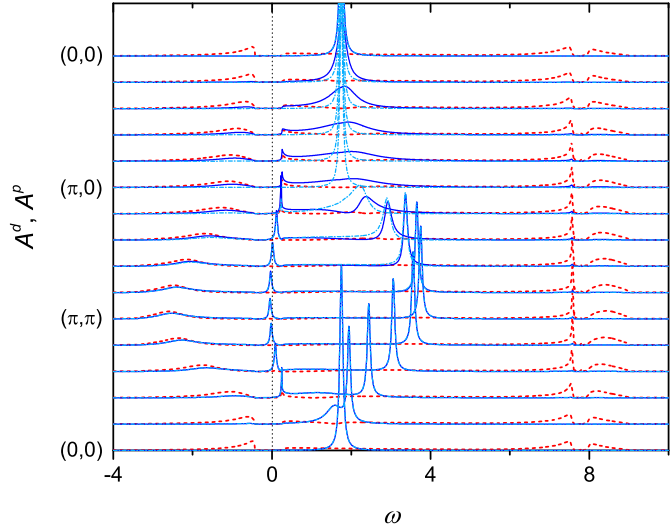}}}
\caption{Spectral functions calculated from Green's functions $G^d({\bf k}\omega)$ (red dashed lines), $G^p({\bf xxk}\omega)$ (blue solid lines) and $G^p({\bf yyk}\omega)$ (cyan dash-dotted lines) along the symmetry lines of the Brillouin zone. Parameters are the same as in Fig.~\protect\ref{Fig2}.} \label{Fig3}
\end{figure}
Spectral functions calculated from copper and oxygen Green's functions (\ref{Gdk}) and (\ref{Gpk2}) are shown in Fig.~\ref{Fig3} for momenta along the symmetry lines of the Brillouin zone. Peaks with predominant oxygen contribution in the frequency range from 2 to 4 belong to an antibonding band, which dispersion owes mainly to a hybridization of symmetric oxygen states with copper states. Except a sharp and intensive peak in the spectrum ${\bf k}=(0,0)$, only few broad maxima [e.g., at ${\bf k}=(0.2\pi,0.2\pi)$] and extended tails in the range from 0 to 2 represent a nonbonding band formed by antisymmetric oxygen states, which do not directly hybridize with copper states and which dispersion is connected mainly with oxygen-oxygen hopping $t_{pp}$. In this frequency range the spectral intensity of this band is nearly completely suppressed by the second term in (\ref{Gpk2}). It can be also concluded from Fig.~\ref{Fig3} that the pseudogap is important for the observation of the Zhang-Rice states -- their peaks are sharp and intensive when they fall into the pseudogap, and these maxima broaden out and blend into the background outside of the pseudogap. This is the reason of the strong sensitivity of Zhang-Rice peaks to parameters, since their location, determined by zeros of the denominator of equation (\ref{Gdk}), as well as the pseudogap location and width depend on these parameters. In particular, an increase of the $d$-$p$ promotion energy $\Delta$ leads to a depletion of the peaks.

The DOS and spectral functions shown in Figs.~\ref{Fig2} and \ref{Fig3} contain the same main features as results of calculations performed by other methods.\cite{Macridin,Weber,Medici,Arrigoni} Our obtained DOS of copper states is similar in shape and has approximately the same relative intensities of maxima as the DOS in Fig.~3 of Ref.~2 (where, however, the upper Hubbard subband is not shown). Also the locations of bands are close to those found in Ref.~5 for the hole-doped case. The exception is the upper Hubbard subband, which is supposed to lie in the range from 2 to 4 (discussing results of other approaches we transform them to the hole picture). If this were the case, the distance between lower and upper Hubbard subbands would be tangibly smaller than $U$, which is hardly possible. Most likely in Ref.~5 the frequency range was chosen too narrow, and it does not contain the upper subband. Dispersions of other bands in Ref.~5 are similar to ours, except for the Zhang-Rice band, which is strongly influenced by the antiferromagnetic ordering taken into account in that work. The agreement with results obtained in the dynamic mean-field approximation is worse, which may be partly connected with differences in parameters. In Ref.~3 the intensity of the Zhang-Rice band is much higher than in our results. In Ref.~4 the intensity of these states is lower, however, the band and the Hubbard subbands have some structure, which is lacking in our spectra.

\section{Concluding remarks}
In this article, the Emery model of Cu-O planes of cuprate high-$T_c$ superconductors was considered using the strong coupling diagram technique. General formulas for calculating hole Green's functions were derived. Using two lowest order terms in the irreducible part and the partial summation, densities of states and spectral functions were calculated for parameters corresponding to hole-doped cuprates. Main features of obtained spectra are similar to those found by other, computationally more complicated methods. In some cases, for close parameters there is agreement in locations, dispersions and relative intensities of bands. The obtained spectra contain Hubbard subbands, lower of which is strongly hybridized with oxygen states for considered parameters, nonbonding and antibonding bands of predominantly oxygen nature, located in the central part of the spectrum, and Zhang-Rice states of mixed nature between the nonbonding band and the lower Hubbard subband. The essential spectral features obtained in our calculations are the spectral pseudogaps. The lower of them is important for the observation of the Zhang-Rice states and explains their sensitivity to parameters. It should be emphasized that the results were obtained in comparatively simple calculations, that gives grounds to suppose that the strong coupling diagram technique may be useful for investigating other generalization of the Hubbard model.

\section*{Acknowledgments}
This work was supported by the research project IUT2-27 and the Estonian Scientific Foundation (grant ETF-9371).

\end{document}